\documentstyle[aps,prl,floats,graphicx,twocolumn]{revtex}

\title{
Point defects on III-V semiconductor surfaces}

\author{
{G.~Schwarz}, 
{J.~Neugebauer}, and 
{M.~Scheffler}
}

\address{
{Fritz-Haber-Institut der Max-Planck-Gesellschaft},\\ 
Faradayweg 4-6, 14195 Berlin-Dahlem, Germany
}

\begin{document}
\draft
\twocolumn[\hsize\textwidth\columnwidth\hsize\csname@twocolumnfalse\endcsname     

\maketitle

\begin{abstract}
The basic properties of  point defects 
(atomic geo\-metry, the position of charge-transfer levels, and formation energies)
on the (110) surface of GaAs, GaP, and InP have been calculated employing
density-functional theory. 
Based on these results we discuss the electronic properties of surface defects,
defect segregation, and compensation. 

\noindent
{
\sf{Sub. to the Proc. of the XXV ICPS, Osaka (Springer, Berlin/Heidelberg, 2000). \\
\copyright{} Springer, Berlin/Heidelberg, 2000.}
}
\end{abstract}

\vskip2pc]
\narrowtext 

\section{Introduction}
\label{intro}

The role of intrinsic point defects in determining the
electronic structure of surfaces is a topic that has been debated
extensively.
They may act as compensation centers and may thus be responsible for 
Fermi-level pinning at the surface.
Consequently, electrically active point defects have been commonly assumed to 
determine Schott\-ky-barrier heights on metal-covered
surfaces\,\cite{spic:88,moen:94-1}. 
Point defects might also act as nucleation centers for crystal growth and
thus influence the surface and interface morphology of devices.
Despite the importance of these \mbox{issues} little is known
about the basic properties of surface point defects, such as atomic geometry,
position of charge-transfer levels, and equilibrium concentration.
In the following we will describe how first-principles calculations can be
employed to identify the basic properties of surface point defects.
Calculations have been performed for GaAs, GaP, and InP.
In the present paper we will limit ourselves mainly to GaAs as a model
system. 
Results for vacancies on GaP\,\cite{schw:97} and InP\,\cite{schwarz:00-1} were
published recently. 

\section{Computational Method}
\label{sec:1}

\begin{figure}[t]
\newlength{\figtwowidth}
\setlength{\figtwowidth}{0.4\hsize}
\centerline{
\begin{tabular}{ll}
\multicolumn{1}{c}{
\includegraphics[width=\figtwowidth]{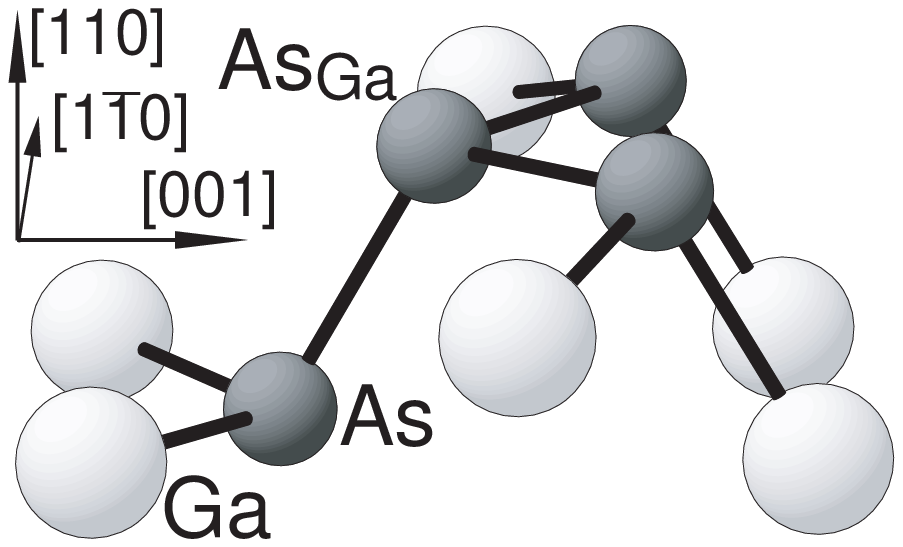}
}
 &
\multicolumn{1}{c}{
\includegraphics[width=\figtwowidth]{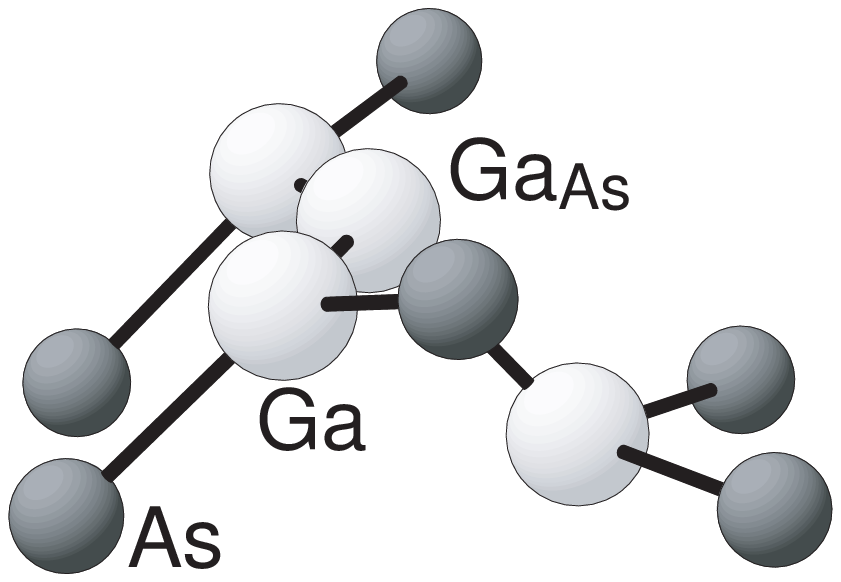}
}
 \\
a) & b) \\ 
\multicolumn{1}{c}{
\includegraphics[width=\figtwowidth]{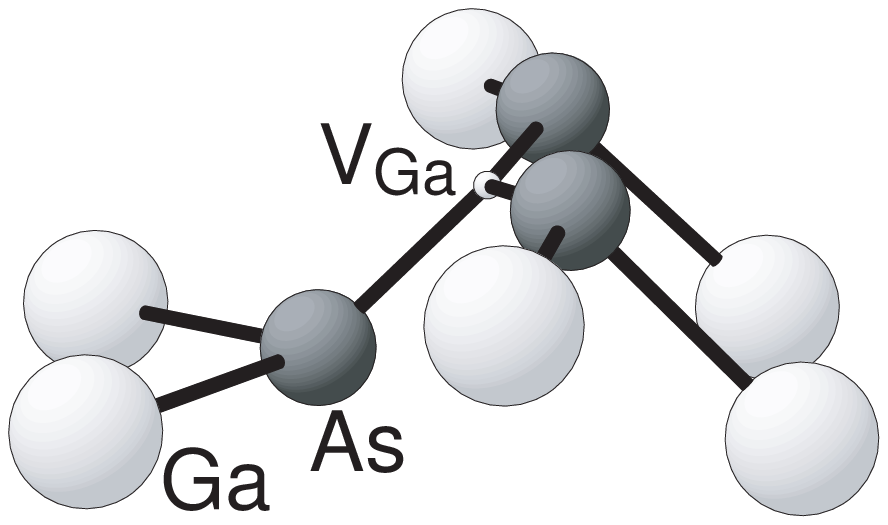}
}
 &
\multicolumn{1}{c}{
\includegraphics[width=\figtwowidth]{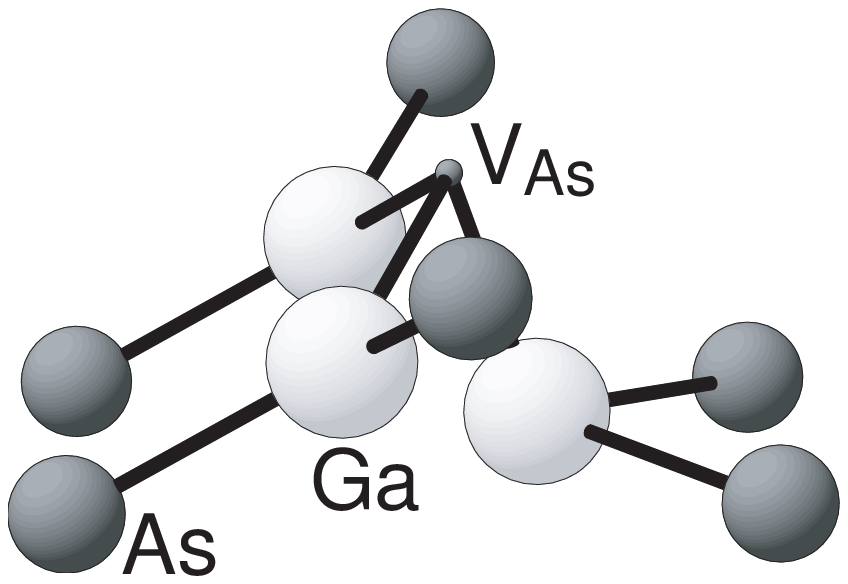}
}
 \\
c) & d) \\
\multicolumn{1}{c}{
\includegraphics[width=\figtwowidth]{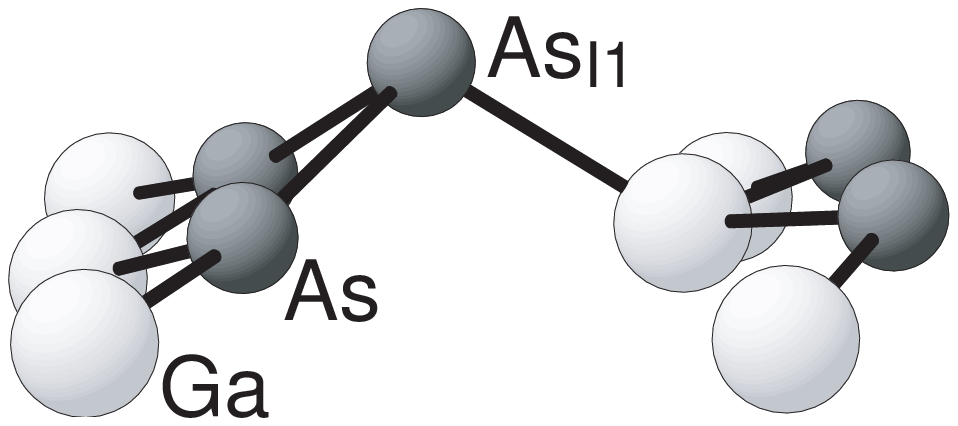}
}
 &
\multicolumn{1}{c}{
\includegraphics[width=\figtwowidth]{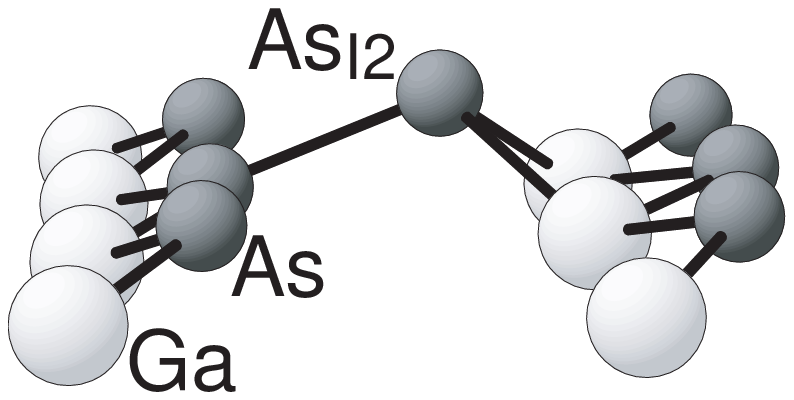}
}
 \\
e) & f) 
\end{tabular}
}
\caption{Atomic relaxations of surface point defects on GaAs(110): antisites
 (a, b), vacancies (c, d), and adatoms (e, f).}
\label{fig:1}       
\end{figure}

In order to investigate the above noted  questions, we performed
density-functional theory (DFT) calculations employing the local-density
approximation (LDA) for exchange and correlation. Details about the method and
extensive convergence tests can be found in Refs.\,\cite{schw:97,fhi96md2}. 
Surface point defects were calculated with six-layer thick slabs.
The (110) surface had a ($2 \times 4$) periodicity
in the [$001$] and [$1\bar{1}0$] direction, respectively (for the defi\-nition of
lattice orientations see the coordinate system in Fig.\,\ref{fig:1}a).
Test calculations showed that these supercells are large enough to calculate
relaxation geometries, electronic structure, and formation energies of point
defects accurately\,\cite{foot1}.

Using the total energy $E$ as calculated for the defect and the surface
systems we obtain the defect-formation energies $E_{\rm f}$ as a function of 
the cation or anion chemical potential and the position of the
surface Fermi level $E_{\rm Fermi}$\,\cite{scheff:93}.
Here we chose the anion chemical potential $\mu_{\rm As}$.
To give an example, the formation energy of an arsenic surface vacancy in the
charge state $q$, $V_{\rm As}^q$, is: 

\begin{equation}\label{equ:1}
 E_{\rm f} (V_{\rm As}^q) = E (V_{\rm As}^q) - E ({\rm slab}) + \mu_{\rm As} + q\, E_{\rm Fermi}\ .
\end{equation}

\noindent
Here $E (V_{\rm As}^q)$ is the total energy of the system with and $E ({\rm slab})$
that of the system without a vacancy. 
The chemical potential of $\rm As$ is controlled by the $\rm As$ partial 
pressure and temperature. 
It can be varied from gallium-rich ($\mu_{\rm As} = -{\rm \Delta} H_f$) to
arsenic-rich conditions ($\mu_{\rm As} = 0$)\,\cite{scheff:93}, where ${\rm
  \Delta} H_f= 0.7$\,eV is the heat of formation of GaAs.
In Fig.\,\ref{fig:2}, $E_{\rm f}$ is plotted as a function of $E_{\rm Fermi}$.
For each point defect and $E_{\rm Fermi}$ only the charge state with mini\-mum
formation energy is shown. 
According to Eq.\,(\ref{equ:1}) the charge state of a defect is given by the
slope of the curve. 
The points where the curve changes it's slope define the positions of $E_{\rm
  Fermi}$ at which the defect changes it's charge state, i.e. the
charge-transfer levels of the defect.

\section{Atomic and electronic structure of point defects}
\label{sec:2}

For several reasons the (110) surface of III-V semiconductors is ideally
suited for the study of intrinsic point defects.
(i) It is the cleavage plane of these materials and high quality surfaces
with a low concentration of steps and defects can be prepared by cleaving
wafers.
Scanning tunneling microscope (STM) experiments achieving atomic
resolution on these surfaces showed point defects such as vacancies and
adatoms in a detailed 
manner\,\mbox{\cite{cox:90-2,gwo:93,leng:94-1,ebert:94-1}.}
(ii) The surface does not reconstruct but undergoes a relaxation that is
characterized by a buckling of surface anions and cations. We calculate the
buckling angle to be $29^\circ$ for GaAs in good agreement with DFT
calculations\,\cite{alv:91} as well as experimental low-energy electron
diffraction analysis\,\cite{duke:85}. 
The point-group symmetry of the surface is $C_{\rm 1h}$ with a single
mirror plane pointing in the [$1\bar{1}0$] direction.
(iii) With the exception of GaP the fundamental band gap of the defect-free
surface is free from surface bands. 
Therefore the presence of point defects with energy levels in the band gap
might notably  affect the electronic properties of the surface.

\subsection{Antisite defects}
\label{subsec:21}

Our calculations show that
arsenic antisite $\rm As_{\rm Ga}$ defects in the surface layer of GaAs show a
rather large displacement of $0.5$\,\AA{} in the  [$00\bar{1}$] and  [$110$]
directions (Fig.\ref{fig:1}a).  
Thus the defect atom is nearly in the same plane as the surface anions: The
large surface buckling between anions and cations of the defect-free
surface is {\em locally} lifted.
The nearest neighbor atoms of  $\rm As_{\rm Ga}$  and all other atoms in the
cell show 
only small relaxations below $0.2$\,\AA{} and $0.05$\,\AA{}, respectively.
This rather localized relaxation pattern is a typical feature of all surface
point defects investigated. 
The atomic structure of the defect-free surface remains nearly unperturbed
beyond second nearest neighbor atoms of surface point defects.
As a further common feature of most point defects we find that the symmetry of
the surface is well conserved by the defect relaxation. 
The only exception where we calculate a nonsymmetric relaxation to have a 
pre\-ferable formation energy is the positively charged anion vacancy which
will be discussed in subsection\,\ref{subsec:22}.  

The atomic relaxation of the gallium surface antisite $\rm Ga_{\rm As}$ is
characterized by a downward and inward movement of the defect with respect to
the surface arsenic atoms. 
The equilibrium position is shifted by $0.5$\,\AA{} and $-0.3$\,\AA{} from the
position of the surface anions in the  [$00\bar{1}$] and [$110$] direction.
The bond length to the neighboring Ga surface atoms is shortened by $3$\,\% 
compared to the defect-free surface. 

Surface antisites on both sublattices do not show any charge-transfer levels
inside the band gap. 
The calculation of the formation energies of the defects reveals that the 
neutral charge state is always preferable energetically.
This is in contrast to antisites in the bulk which we calculate as
double acceptors ($\rm Ga_{\rm As}$) and \mbox{double} donors  ($\rm As_{\rm Ga}$) in
agreement with previous theoretical 
studies \mbox{\cite{bar:85,scheff:88-2,zhang:91}}.
We therefore predict that surface antisites will remain charge neutral on
intrinsic as well as on $p$- or $n$-type doped material and that they do not
alter the {\em electronic} properties of the (110) surface.
We note that the antisites are electrically inactive at the surface only. 
Once incorporated into the bulk (e.g. at low temperatures where the surface
concentration of defects will be frozen during growth) they become
electrically active. 
This change in character is rather abrupt as we find anti\-sites already in the
second layer to be similar to bulk ones in terms of formation energy and
electronic structure in agreement with other calculations\,\cite{jan:95}.

\subsection{Vacancies}
\label{subsec:22}

As illustrated in Fig.\,\ref{fig:1} a vacancy might be derived from an antisite
by removing the defect atom. 
By doing so the atoms surrounding the defect change their equilibrium
positions significantly, and we find a very distinct relaxation pattern. 
The relaxations of the neutral surface vacancies on GaAs are shown in
Fig.\,\ref{fig:1}c,d. 
On both sublattices they are characterized by a comparatively large relaxation
of the surface nearest neighbor atoms into the surface
by $0.5$\,\AA{} ($V_{\rm Ga}$) and $0.3$\,\AA{} ($V_{\rm As}$).
The neighboring atom in the second layer also relaxes into the void of the
vacancy by $0.5$\,\AA{} for both defects.
The $C_{\rm 1h}$ symmetry of the defect-free surface 
is conserved by the relaxation of the neutral vacancies. 
The anion vacancy in the positive charge state, however, prefers a
nonsymmetric, rebonded configuration where one of the surface cations
approaches the cation in the second layer.
The energy gain of this configuration compared to the symmetric relaxation
is $0.10$\,eV on GaP, $0.07$\,eV on InP\,\cite{schwarz:00-1}, and $0.17$\,eV
on GaAs.  
The latter value as well as the ground-state geometry is in good agreement
with the result of $0.16$\,eV by Zhang and Zunger\,\cite{zhang:96-1} for
$V_{\rm As}$ on GaAs. 

Based on our calculations we find that surface vacancies are electrically active.
They have an amphoteric character on all three materials investigated,
i.e., they are  positively charged on $p$- and negatively charged on
$n$-type material. 
$V_{\rm Ga}$ on GaAs has two charge-transfer levels $E_{\rm TL} (+,0)$ and
$E_{\rm TL} 
(0,-)$ in the lower half of the band gap while $V_{\rm As}$ is a negative U
center with $E_{\rm TL} (+,-)$ in the lower half of the band gap  as shown in
Fig.\,\ref{fig:2}. 
We can therefore conclude that vacancies are efficient compensation centers
for $p$- and $n$-type conditions.

\subsection{Adatoms}
\label{subsec:23}
\begin{figure}
\newlength{\figthreewidth}
\setlength{\figthreewidth}{0.9\hsize}
\includegraphics[width=\figthreewidth]{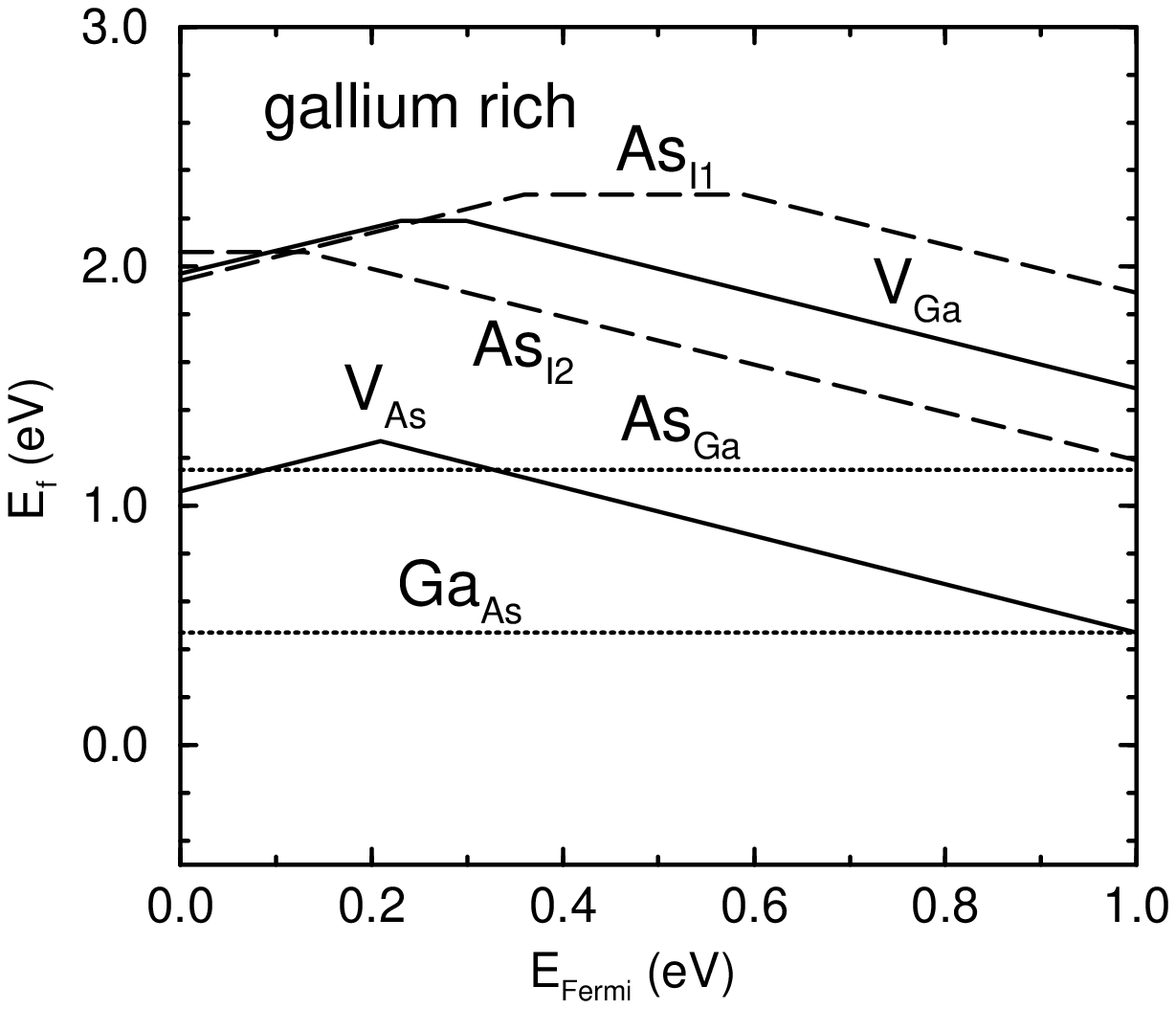}\\
a)\\
\includegraphics[width=\figthreewidth]{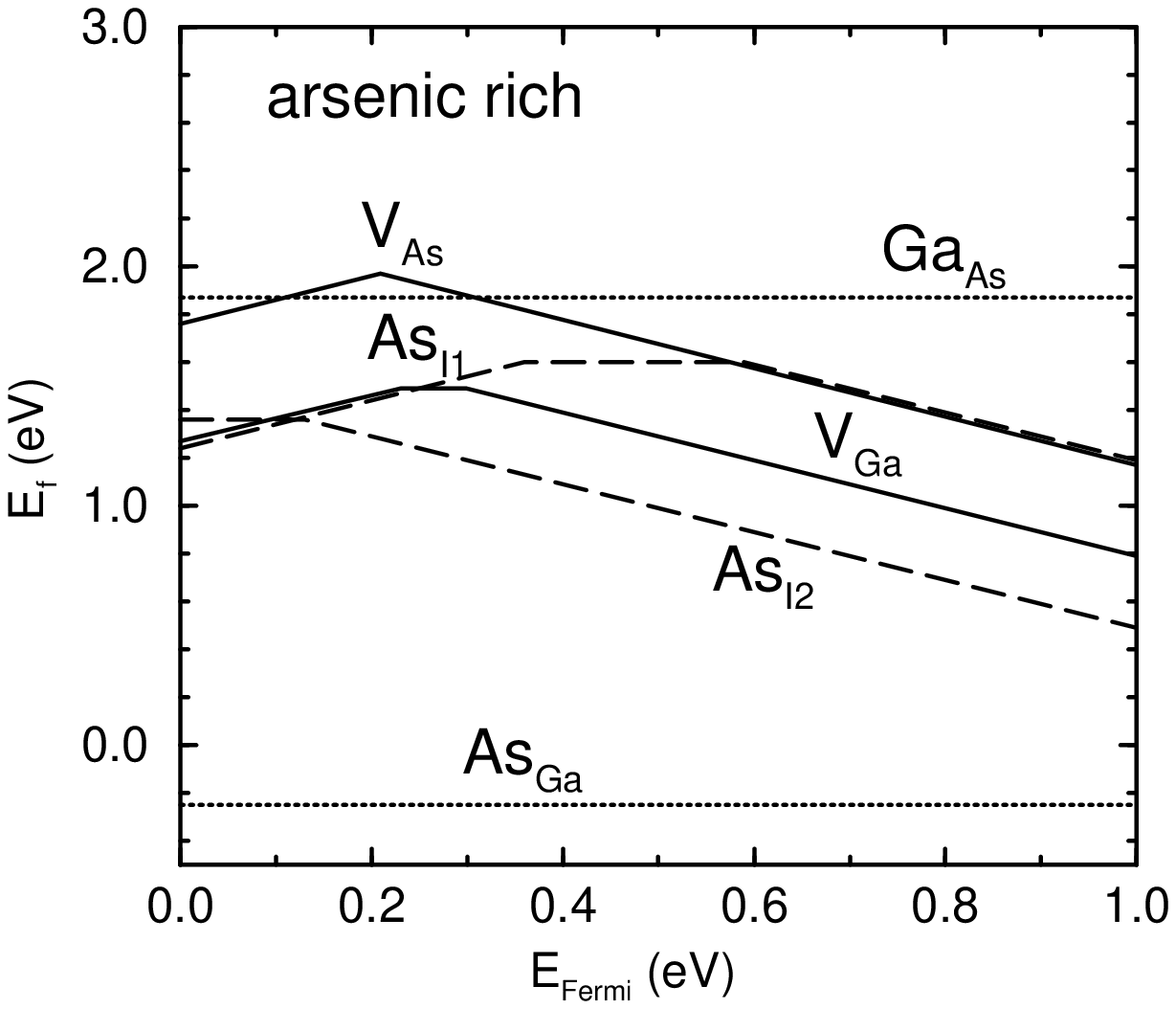}\\
b)
\caption{Formation energies of surface point defects on GaAs(110) as a
  function of the surface Fermi level $E_{\rm Fermi}$. (a) gallium-rich
  conditions with \mbox{$\mu_{\rm As} = -0.7$\,eV} as compared to bulk arsenic. 
  (b) arsenic-rich conditions with \mbox{$\mu_{\rm As} = 0$\,eV}.
  Changing to  arsenic-rich conditions vacancies on the
  cation and anion sublattice (solid lines) change formation energy by $-0.7$ and
  $+0.7$\,eV, respectively. 
  Antisites (dotted lines) change by $+1.4$ ($\rm Ga_{\rm As}$) and  $-1.4$\,eV ($\rm As_{\rm
  Ga}$) while the adatoms (dashed lines) are more preferable by  $-0.7$\,eV.
}
\label{fig:2}       
\end{figure}

As a last type of surface point defects we have calculated anion adatoms. As
illustrated in Fig.\,\ref{fig:1} we find two possible configurations: one in
which the adatom is bonded to the surface anions ($\rm As_{\rm I1}$,
Fig.\,\ref{fig:1}a) and one with bonds to the surface cations ($\rm As_{\rm I2}$,
Fig.\,\ref{fig:1}b).
In the configuration $\rm As_{\rm I1}$ the distance between the adatom and the
arsenic surface atoms is $2.61$\,\AA{} in the neutral charge state while for
$\rm As_{\rm I2}$ the distance to the surface gallium atoms is  $2.64$\,\AA{}.
The adatom is located $1.1$\,\AA{} ($\rm As_{\rm I1}$) and $0.8$\,\AA{} ($\rm
As_{\rm I2}$) above the surface.
The defect $\rm As_{\rm I1}$ bound to surface arsenic has an electronic
structure similar to the surface vacancies showing two charge-transfer \mbox{levels}
$E_{\rm TL} (+,0)$ and $E_{\rm TL} (0,-)$ (Fig.\,\ref{fig:2}). 
In contrast, $\rm As_{\rm I2}$, which bonds to the surface cations, is predicted
to be nega\-tively charged for all positions of the Fermi level except for
extreme $p$-type conditions with $E_{\rm Fermi}$ below $0.1$\,eV, where  the
neutral charge state is found to be stable.

Unlike previous first-principles calculations by Yi {\it et
  al.}\,\cite{yel:95} the adatom $\rm As_{\rm I2}$ is determined as the
preferred adsorption site over a wide range of the surface Fermi level
(Fig.\,\ref{fig:2}).  
Only for $p$-type conditions near the valence-band maximum $\rm As_{\rm I1}$
in the singly positive charge state has a lower formation energy than $\rm
As_{\rm I2}$. 
For the neutral charge states the difference in formation energy is $0.2$\,eV
while Yi {\it et al.}\,\cite{yel:95} report a difference of $-0.1$\,eV.
This difference might be due to the limited size of the surface unit cell ($2 
\times 2$) and the use of the $\rm \Gamma$ point only for Brillouin zone
integration as employed in Ref.\,\cite{yel:95}. 

%

 
\section{Formation energies and defect concentrations}
\label{sec3}

Based on the formation energies as shown in Fig.\,\ref{fig:2} we identify the
dominant point defects on the surface
 and how the concentration depends on the anion chemical potential and doping
conditions. 
Under gallium-rich conditions the gallium antisite $\rm Ga_{\rm As}$ is lowest
in formation energy (Fig.\,\ref{fig:2}a). It is, however, not electrically
active and thus can be excluded as a possible dopant or compensation center.
The dominant point defect on GaAs under gallium-rich conditions is the arsenic
vacancy 
$V_{\rm As}$, positively charged under $p$-type and negatively charged
under $n$-type conditions. 
This conclusion changes if we consider arsenic-rich and $p$-type conditions.
Then $V_{\rm Ga}$ and $\rm As_{\rm I1}$, both in the singly positive charge
state, are almost degenerate in formation energy (Fig.\,\ref{fig:2}b). 
With $E_{\rm Fermi}$ in the upper half of the band gap the negatively charged
$\rm As_{\rm I2}$ is the defect with the lowest formation energy among all
electrically active point defects.
Fig.\,\ref{fig:2}b also shows that the neutral $\rm As_{\rm Ga}$ has a
{\em negative formation energy} under arsenic-rich conditions. 
This indicates that the surface is unstable against the formation of a
complete layer of arsenic. 
Indeed, first-principles calculations of the surface energy show this structure
to be lower in energy than the cleavage surface assuming extreme arsenic-rich conditions\,\cite{moll:96}.

Based on the calculated formation energies  we can immediately obtain the 
equilibrium concentration $C$ of point defects. Neglecting vibrational entropy
contributions this quantity is determined by the number of possible sites
$C_0$ where a defect can be formed, the formation energy $E_{\rm f}$, and
temperature $T$: 

\begin{equation}
 C = C_0 \, e^
     {- \, \frac{E_{\rm f} (\mu_{\rm As},E_{\rm Fermi})}{k_{\rm B}T}}
\ .
\end{equation}

\noindent
Here $k_{\rm B}$ is Boltzmann's constant. 
Using this formalism, assuming room temperature, and 
taking into account the electrically active defect with the lowest $E_{\rm f}$
($0.47$\,eV for $V_{\rm As}$ under gallium-rich conditions with $E_{\rm
  Fermi}$ at the conduction-band minimum) we find a maximum equilibrium
concentration of about $10^6\,cm^{-2}$. 
The concentration decreases if we 
assume a situation where the Fermi level is fully pinned at the
charge-transfer level of the vacancy at $0.2$\,eV above the valence-band
maximum. 
Then the formation energy becomes $1.27$\,eV and the concentration is
negligible ($10^{-9}\,cm^{-2}$).
A simple analysis shows that pinning of the surface Fermi level by defects
requires rather large concentrations of about
$10^{12}\,cm^{-2}$\,\cite{moen:93-1}. 
This conclusion is consistent with com\-bined STM
and photoelectron spectros\-copy measure\-ments on InP(110), which showed
significant surface band bending 
at concentrations of phosphorus surface vacancies higher than $2\times
10^{12}\,cm^{-2}$\,\cite{schwarz:00-1}. 
However, also for the phosphorus surface vacancy we find an equilibrium
concentration (at $430$\,K where the experiment has been performed)
significantly below the experimentally observed defect concentration.
We therefore conclude that the formation of vacancies is driven by kinetic
mechanisms rather than as a result of thermal equilibrium.
A possible kinetic processes on the surface is 
diffusion from the bulk driven by the lower formation energy of surface
vacancies compared to their bulk counterparts.

\section{Summary and conclusions}

Density-functional theory has been used to study point defects on the (110) surface
of GaAs.
We identify anti\-sites on both sublattices to be the defects with the lowest
formation energy.  
They are not electrically active and can therefore be excluded as
compensation centers on this surface. 
Among the electrically active point defects $V_{\rm As}$ is lowest in
formation energy under gallium-rich conditions. 
For arsenic-rich conditions and $n$-type doping $V_{\rm Ga}$ and arsenic
adatoms are predicted to be most important.
The calculated formation energies allow the estimation of the equilibrium
concentrations of surface defects. 
We have shown that the formation energy of all electrically active defects is too
high to allow for {\em equilibrium} concentrations large enough to induce
significant band bending at the surface.
The high concentration of defects reported in recent experiments is thus
related to kinetic effects.

This work was supported by the `Deutsche Forschungs\-gemeinschaft'
through `Sonder\-for\-schungs\-bereich' 296 TP A5.
We like to thank Ph.~Ebert and K.~Horn for stimulating discussions.


\begin{thebibliography}{10}

\bibitem{spic:88}
W. E. Spicer \textit{et~al.}, J. Vac. Sci. Technol. B \textbf{6}, (1988) 1245.

\bibitem{moen:94-1}
W. M\"onch, Surf. Sci. \textbf{299/300},  (1994) 928.

\bibitem{schw:97}
G. Schwarz, A. Kley, J. Neugebauer, and M. Scheffler, Phys. Rev. B \textbf{58},
    (1998) 1392.

\bibitem{schwarz:00-1}
P. Ebert \textit{et~al.}, Phys. Rev. Lett. \textbf{84},    (2000) 5816.

\bibitem{fhi96md2}
M. Bockstedte, A. Kley, J. Neugebauer, and M. Scheff\-ler, Comp. Phys. Commun.
  \textbf{107},    (1997) 187, see also: 
{http://www.fhi-berlin.mpg.de/th/fhimd/}.

\bibitem{foot1}
Gallium surface vacancies on GaAs (110) change formation energies and the
  position of Kohn-Sham eigenstates in the band gap by less than 10\,meV when
  increasing the supercell to ($2 \times 5$) or ($3 \times 4$) periodicity.
  Relaxation pattern coincide in these calculations better than by
  $0.01$\,\AA{}.

\bibitem{scheff:93}
U. Scherz and M. Scheffler,  in \textit{Imperfections in III-V Materials},
  semiconductors and semimetals, edited by E.~R. Weber (Academic Press, New
  York, 1993), Vol.~38, Chap.~1, p.\ 1.

\bibitem{cox:90-2}
G. Cox \textit{et~al.}, Vacuum \textbf{41},    (1990) 591.

\bibitem{gwo:93}
S. Gwo, A.~R. Smith, and C.~K. Shih, J. Vac. Sci. Technol. A \textbf{11},  
  (1993) 1644.

\bibitem{leng:94-1}
G. Lengel, R. Wilkins, and M. Weimer, Phys. Rev. Lett. \textbf{72},    (1994)
836.

\bibitem{ebert:94-1}
P. Ebert, K. Urban, and M.~G. Lagally, Phys. Rev. Lett. \textbf{72},
(1994) 840.

\bibitem{alv:91}
J.~L.~A. Alves, J. Hebenstreit, and M. Scheffler, Phys. Rev. B \textbf{44},  
  (1991) 6188.

\bibitem{duke:85}
C.~B. Duke, C. Mailhiot, and A. Paton, J. Vac. Sci. Technol. B \textbf{3},  
  (1985) 1087.

\bibitem{bar:85}
G.~A. Baraff and M. Schl\"uter, Phys. Rev. Lett. \textbf{55},    (1985) 1327.

\bibitem{scheff:88-2}
J. D\c{a}browski and M. Scheffler, Phys. Rev. Lett. \textbf{60},    (1988) 2183.

\bibitem{zhang:91}
S.~B. Zhang and J.~E. Northrup, Phys. Rev. Lett. \textbf{67},   (1991) 2339.

\bibitem{jan:95}
R.~B. Capaz, K. Cho, and J.~D. Joannopoulos, Phys. Rev. Lett. \textbf{75},  
  (1995) 1811.

\bibitem{zhang:96-1}
S.~B. Zhang and A. Zunger, Phys. Rev. Lett. \textbf{77},    (1996) 119.

\bibitem{yel:95}
J.-Y. Yi \textit{et~al.}, Phys. Rev. B \textbf{52},    (1995) 10733.

\bibitem{moll:96}
N. Moll, A. Kley, E. Pehlke, and M. Scheffler, Phys. Rev. B \textbf{54},  
  (1996) 8844.

\bibitem{moen:93-1}
W. M\"onch, \textit{Semiconductor Surfaces and Interfaces},
  Springer Series in Surface Sciences Vol.~26, edited by G.~Ertl. \textit{et
    al.} (Springer, Berlin/Heidelberg, 1993).

\end{thebibliography}
\end{document}